\documentclass{kapproc}
\usepackage{t1enc}
\usepackage{procps} 
\usepackage[dvips]{graphicx}
\upperandlowercase
\kluwerbib

\begin{document}

\articletitle{SHADES: The Scuba HAlf Degree\,\\ Extragalactic Survey}

\author{James S. Dunlop}
\affil{Institute for Astronomy, University of Edinburgh, UK}
\email{jsd@roe.ac.uk}

\begin{abstract}
SHADES is a new, major, extragalactic sub-mm survey currently being
undertaken with SCUBA on the JCMT. The aim of this survey is to map
0.5 square degrees of sky at a depth sufficient to provide the first,
major ($\simeq 300$ source), unbiased sample of bright ($S_{850}
\simeq 8$ mJy) sub-mm sources. Combined with extensive multi-frequency
supporting observations already in hand, we aim to measure the
redshift distribution, clustering and AGN content of the sub-mm
population. Currently $\simeq 40$\% complete, the survey is due to run
until early 2006. Here I provide some early example results which
demonstrate the potential power of our combined data set, and
highlight a series of forthcoming papers which will present results
based on the current interim sample of $\simeq 130$ $850\mu$m sources
detected within the Lockman Hole and SXDF SHADES survey fields.
\end{abstract}

\section{Survey Rationale}

\begin{figure}[ht]
\centering
   \vspace*{17cm}
   \includegraphics{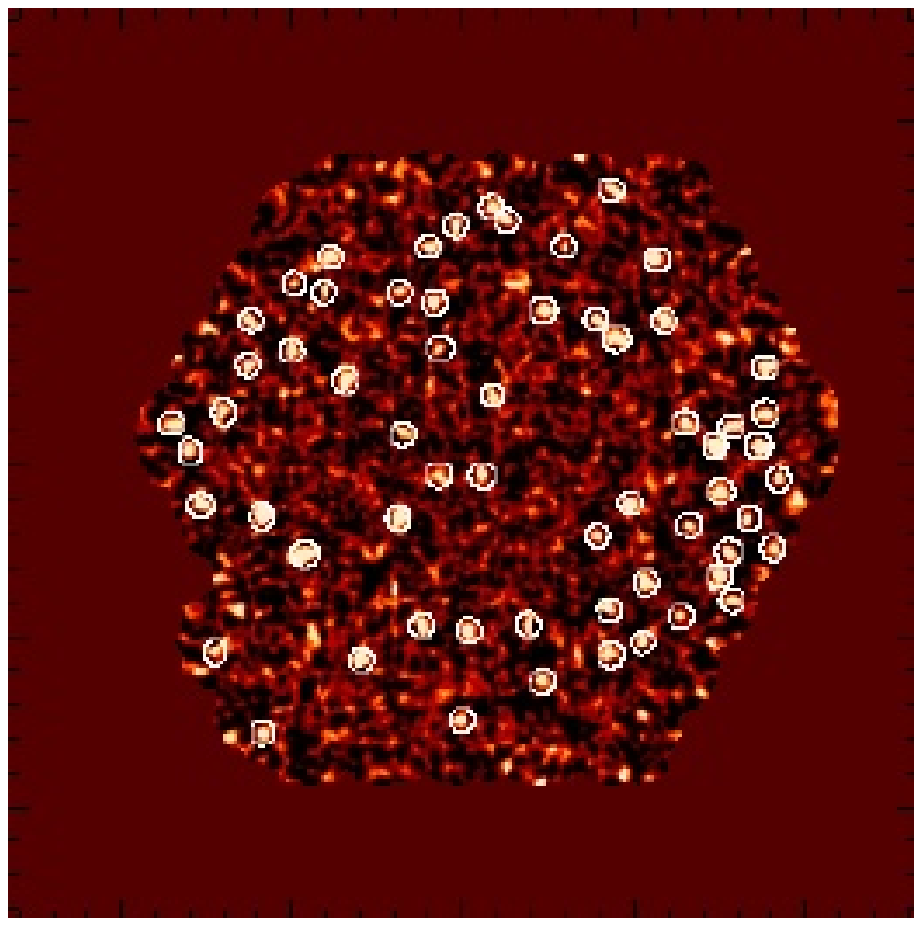}
   \includegraphics{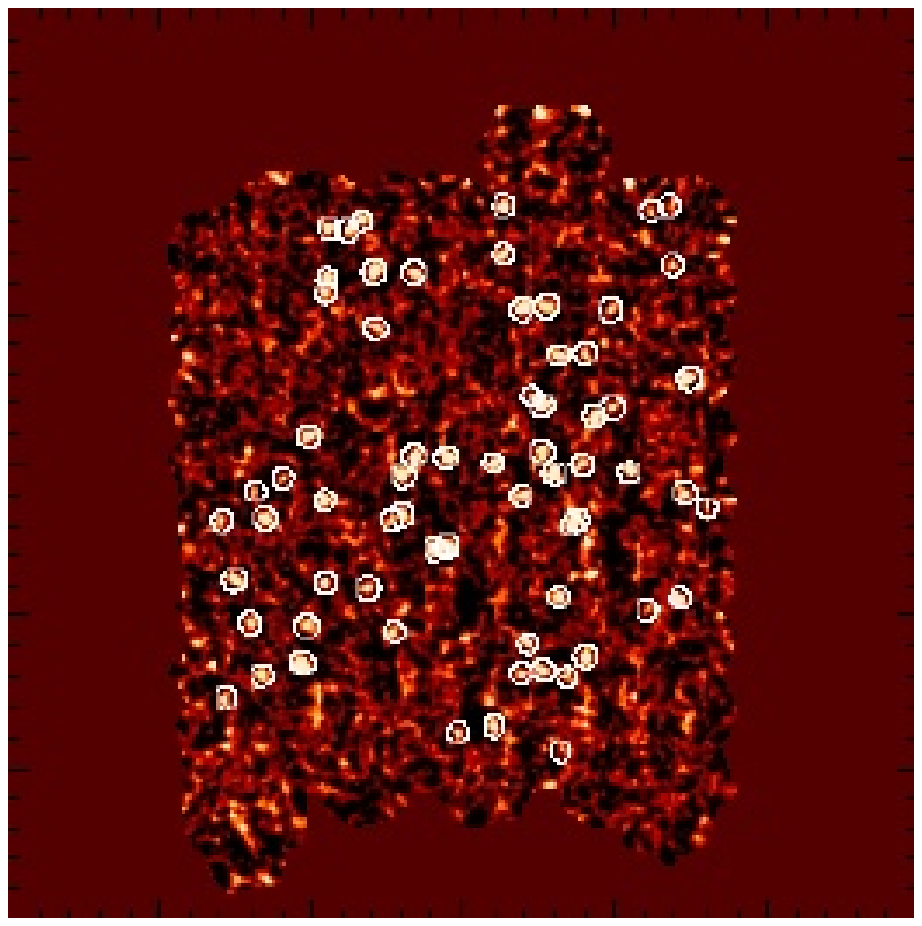}
\caption{Signal-to-Noise images of the current $850\mu$m SHADES images
of the SXDF (upper plot), and Lockman Hole (lower plot). Sources
confirmed by the multiple reductions are circled.}
\end{figure}

The sub-mm galaxy population continues to present a major challenge to
theories of galaxy formation (e.g., Baugh et al. 2004, Somerville
2004), as current semi-analytic models cannot naturally explain the
existence of a substantial population of dust-enshrouded starburst
galaxies at high redshift. However, while now regarded as of key
importance by theorists, the basic properties of sub-mm galaxies are
not, in fact, well defined. Several redshifts have been measured
(Chapman et al. 2003), some masses have been determined from CO
observations (Genzel et al. 2004), and several individual
SCUBA-selected galaxies have been studied in detail (e.g., Smail et
al. 2003).  However, these follow-up studies have had to rely on
small, inhomogeneous, and often deliberately biased (e.g., lensed or
radio pre-selected) samples of sub-mm sources, and until now no
robust, complete, unbiased and statistically significant (i.e., $>
100$ sources) sample of sub-mm sources has been constructed.

SHADES ({\tt http://www.roe.ac.uk/ifa/shades}), the Scuba Half Degree
Extragalactic Survey, was designed to remedy this situation. It aims
to map 0.5 square degrees with SCUBA to an r.m.s.  noise level of
$\simeq 2$ mJy at $850\mu$m. The SHADES consortium consists of nearly
the entire UK sub-mm cosmology community, coupled with a subset of the
BLAST balloon-borne observatory consortium.

The survey has many goals (see Mortier et al. 2004), but the primary
objective is to provide a complete and consistent sample of a few
hundred sources with $S_{850} > 8$ mJy, with sufficient supporting
multi-frequency information to establish the redshift distribution,
clustering properties, and AGN fraction of the bright sub-mm
population. The aim is to provide this information, within the 3-year
lifetime of the survey, by co-ordinating the SCUBA mapping
observations with (i) deep VLA and Merlin radio imaging, (ii) Spitzer
mid-infrared imaging, (iii) far-infrared maps of the same fields made
with BLAST, (iv) optical and near-infrared imaging with UKIRT and the
Subaru telescope, and (v) deep optical spectroscopy with Gemini, Keck
and the VLT.

\section{SCUBA mapping}

SHADES is split between two fields -- the Subaru SXDF field at RA
$02^{\rm h} 18^{\rm m}\\ 00^{\rm s}$, Dec $-05^\circ 00' 00''$ (J2000),
and the Lockman Hole centred on RA $10^{\rm h} 52^{\rm m}\\ 51^{\rm s}$,
Dec $57^\circ 27' 40''$ (J2000), with the goal being to map 0.25
square degrees in each. These two fields were chosen both to provide a
spread in RA (to allow observation with the JCMT throughout the
majority of the year), and because each field offered unique
advantages in terms of low Galactic cirrus emission (crucial for BLAST
and Spitzer observations) and existing/guaranteed supporting data at
other wavelengths.

SHADES SCUBA observations commenced at the JCMT in December 2002.
Full details on the observing technique can be found in Mortier et
al. (2004).

\begin{figure}[ht]

\centering
   \vspace*{15cm}
   \includegraphics{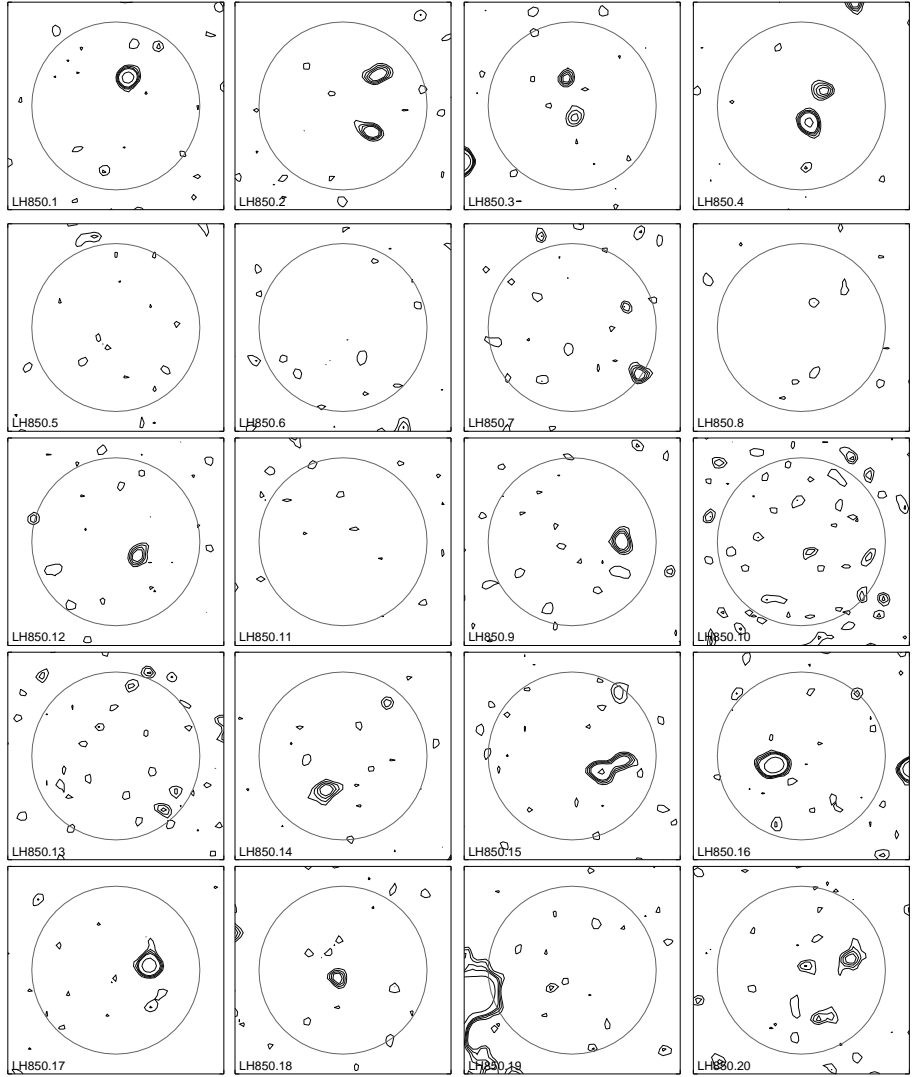}
\caption{VLA postage stamps centred on the positions of the top 20
SHADES sources in the Lockman Hole field. 70\% of these 20 sources
have a VLA counterpart within the illustrated search radius of 8
arcsec, and consideration of the lower-resolution B+C array data alone
raises this figure to 80\% (see Fig. 3)}
\end{figure}

SCUBA Signal-to-Noise maps for the SXDF and Lockman SHADES fields
obtained by Spring 2004 are shown in Fig. 1.  The total area covered
by these two maps is 700 square arcmin (402 square arcmin in the
Lockman Hole, 294 square arcmin in the SXDF), of which an effective
area of $\simeq 650$ square arcmin has complete coverage.

The survey is therefore $\simeq 40\%$ complete, and to date has
yielded a total of 130 sources, whose reality is confirmed by 4
independent reductions undertaken within the consortium.  (61 sources
in the SXDF image (24 at $> 4 \sigma$, 53 at $> 3.5 \sigma$) and 69
sources in the Lockman Hole image (22 at $> 4 \sigma$, 47 at $> 3.5
\sigma$)).  Based on this interim reduction, we predict a final source
list of $\simeq 300$ sources.

\begin{figure}[ht]
\vspace*{7cm}
\centering
   \includegraphics{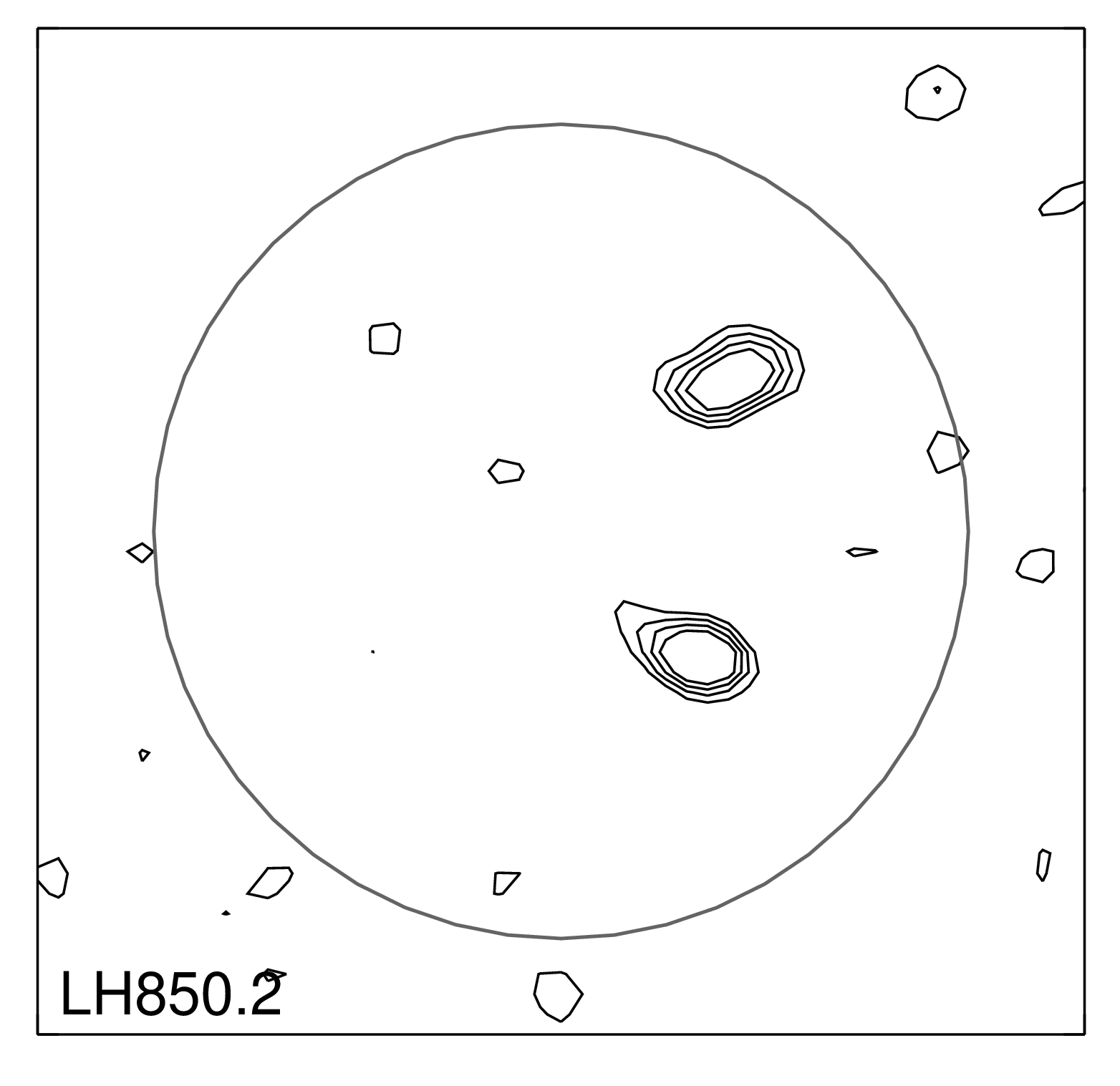}
   \includegraphics{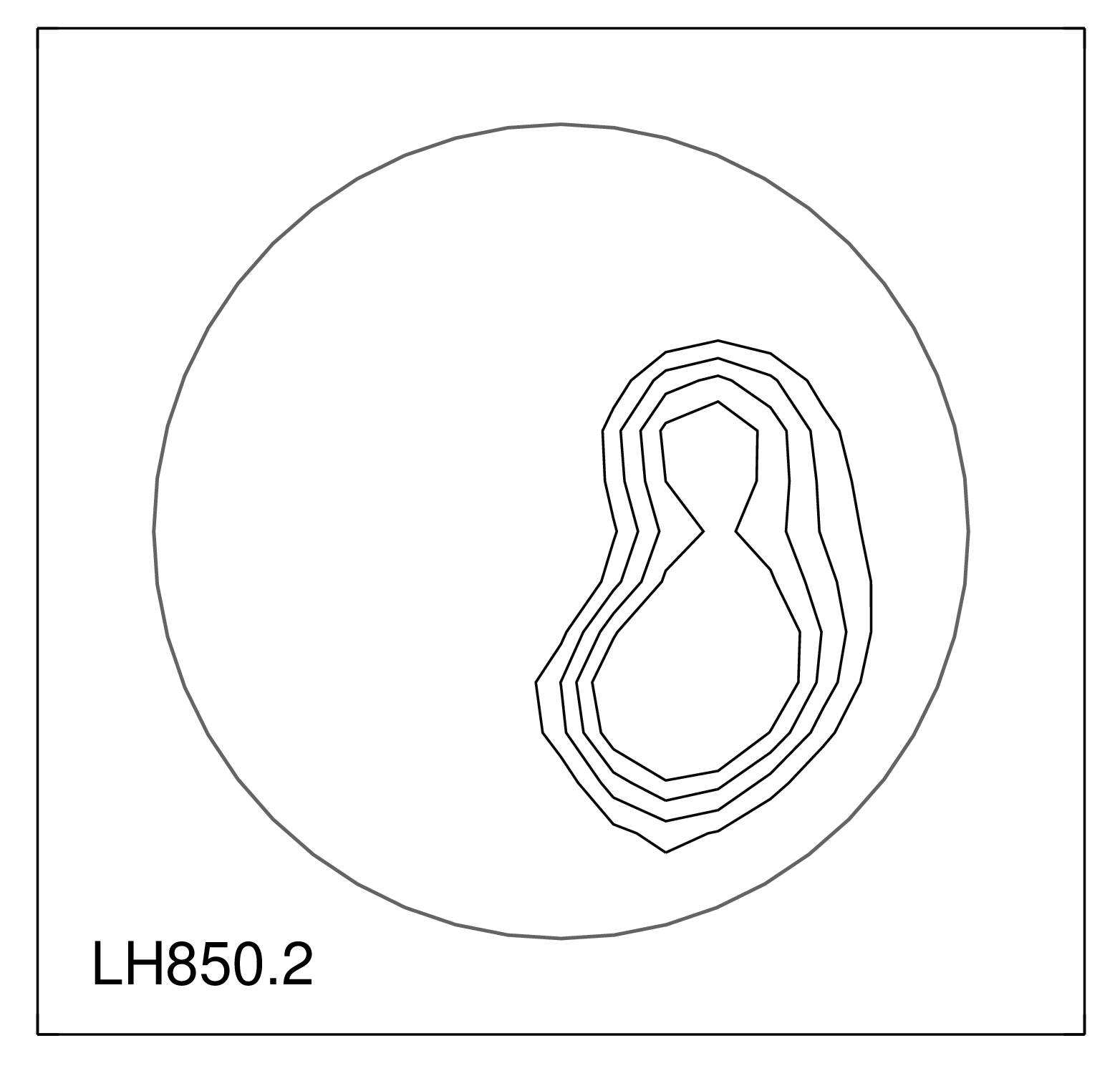}
   \includegraphics{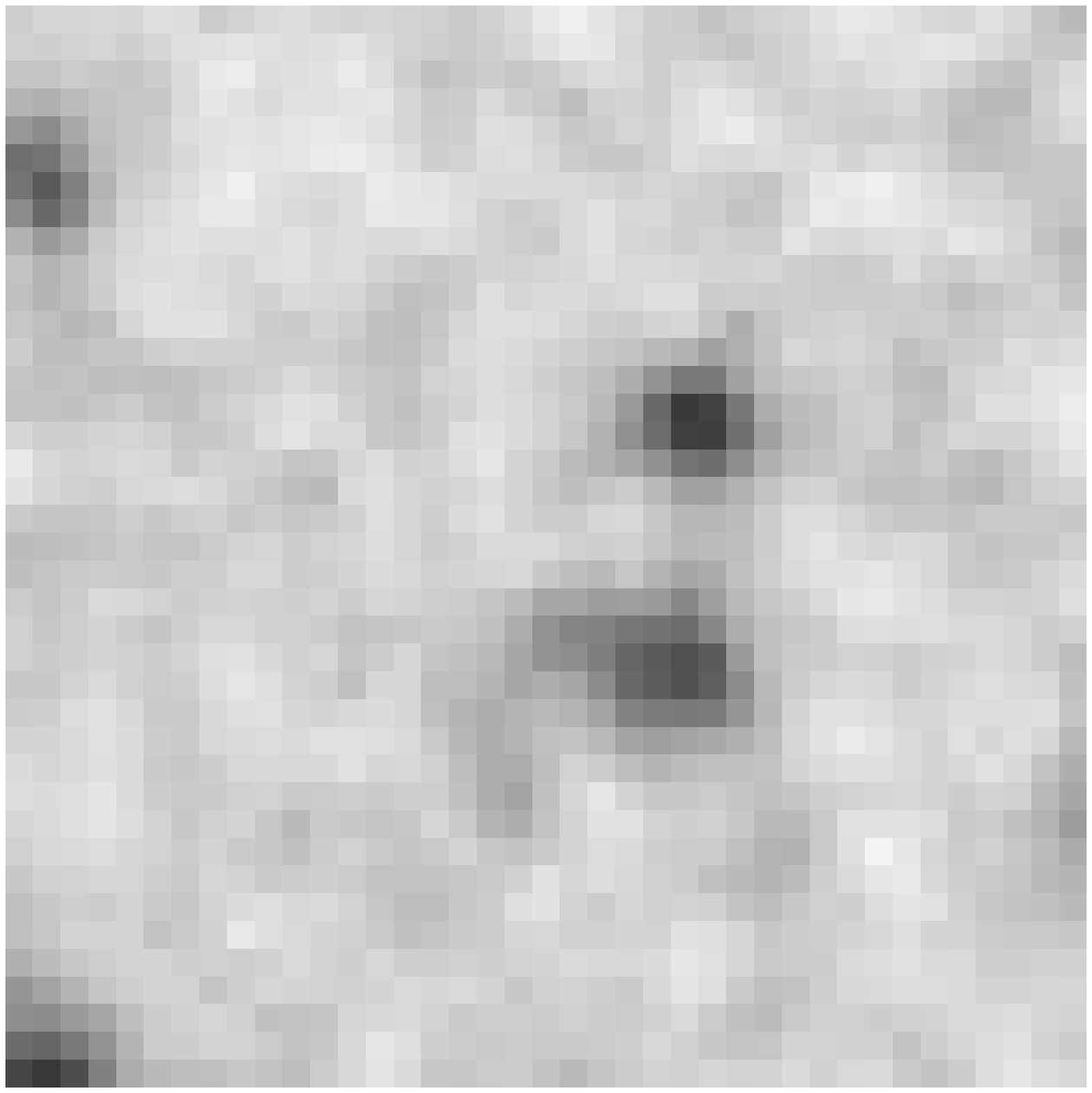}
   \includegraphics{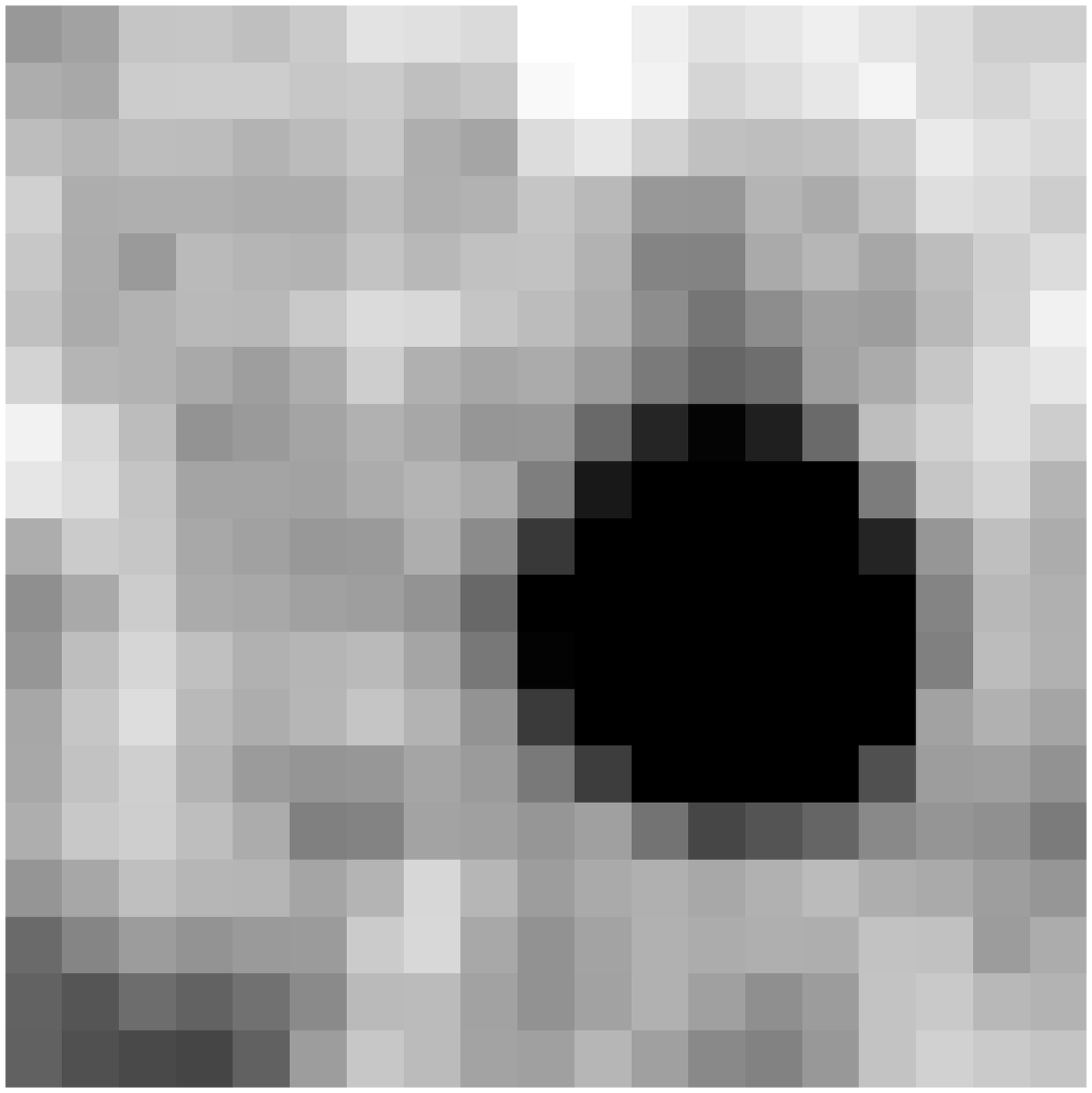}

   \includegraphics{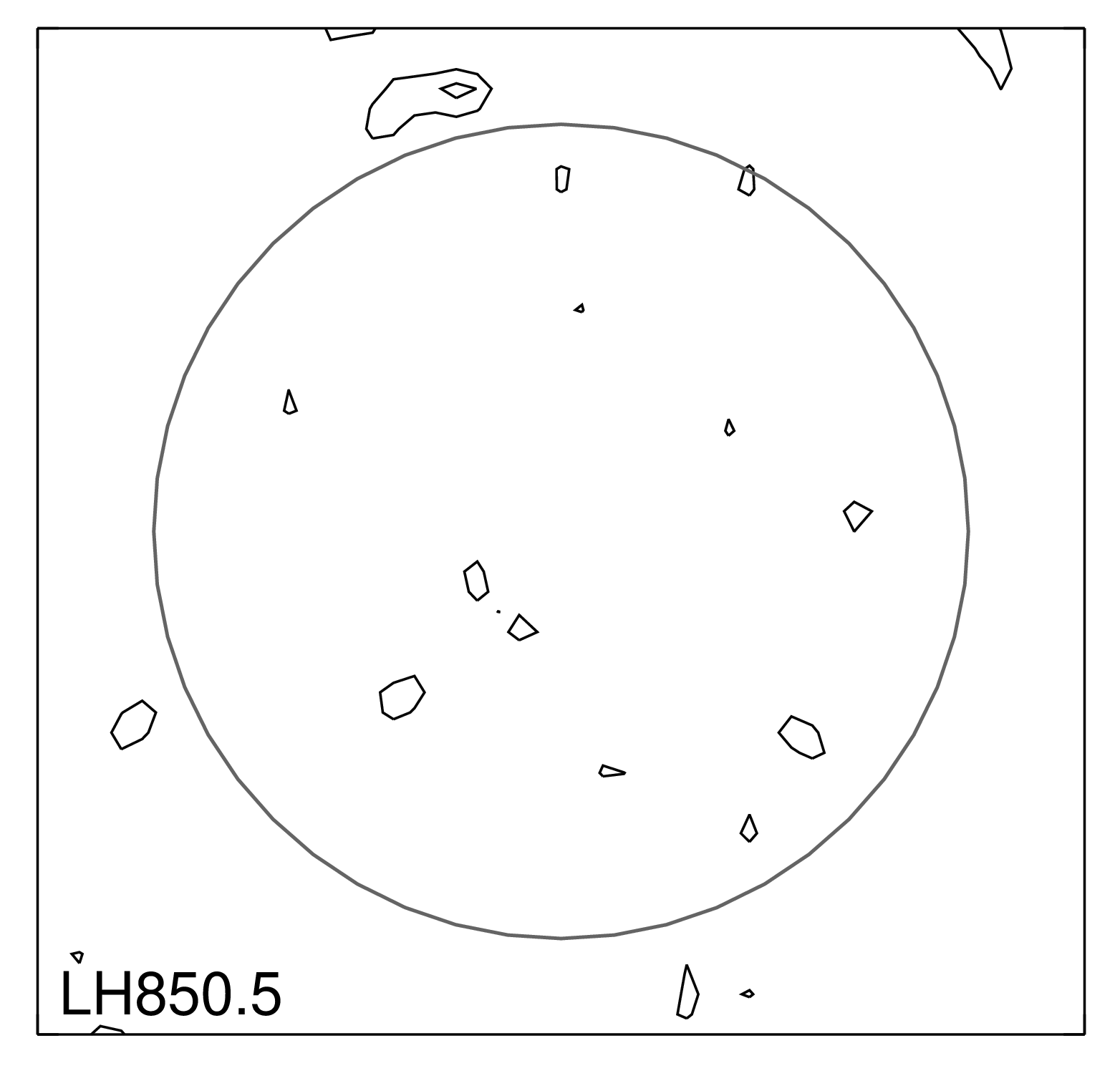}
   \includegraphics{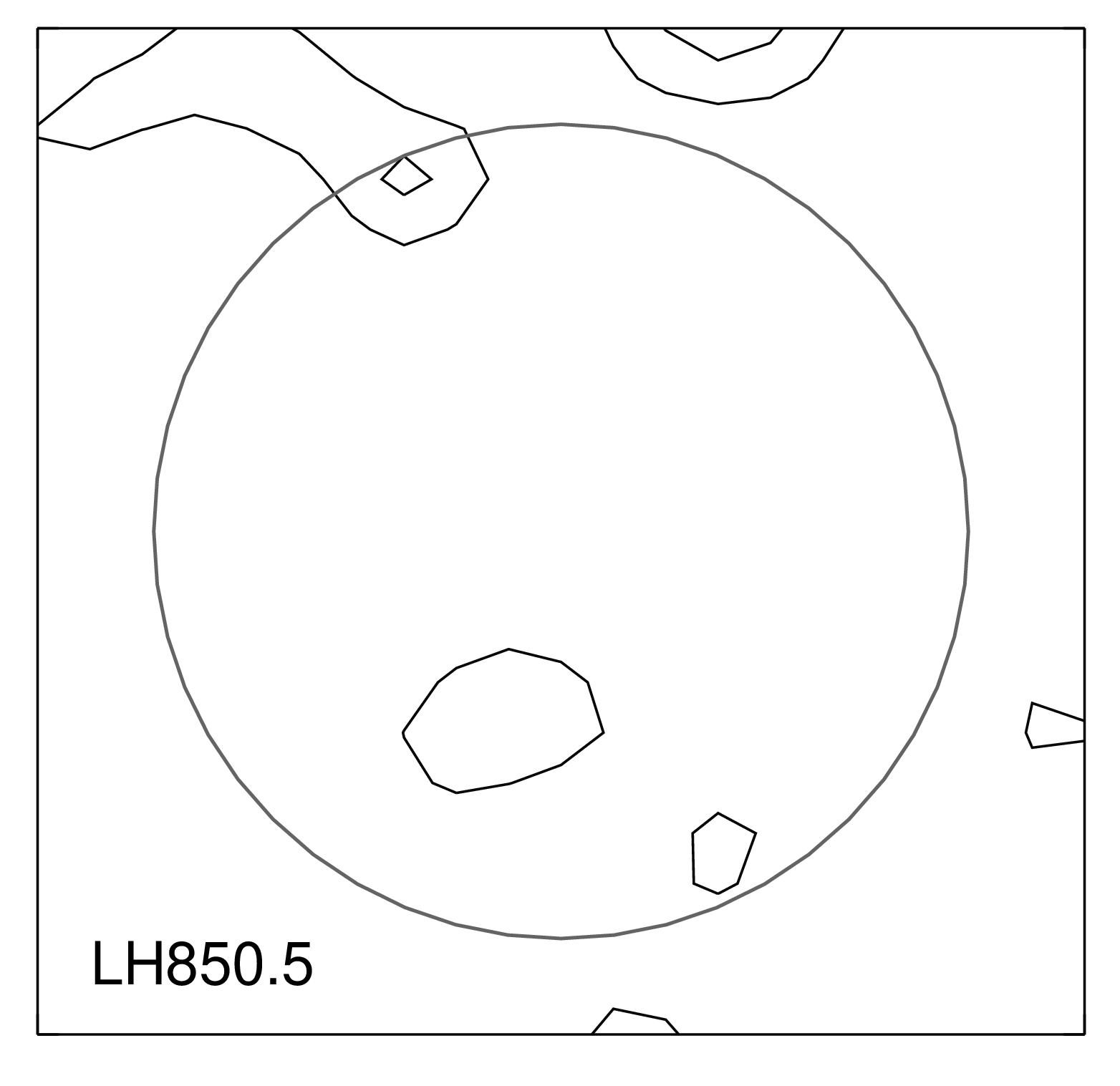}
   \includegraphics{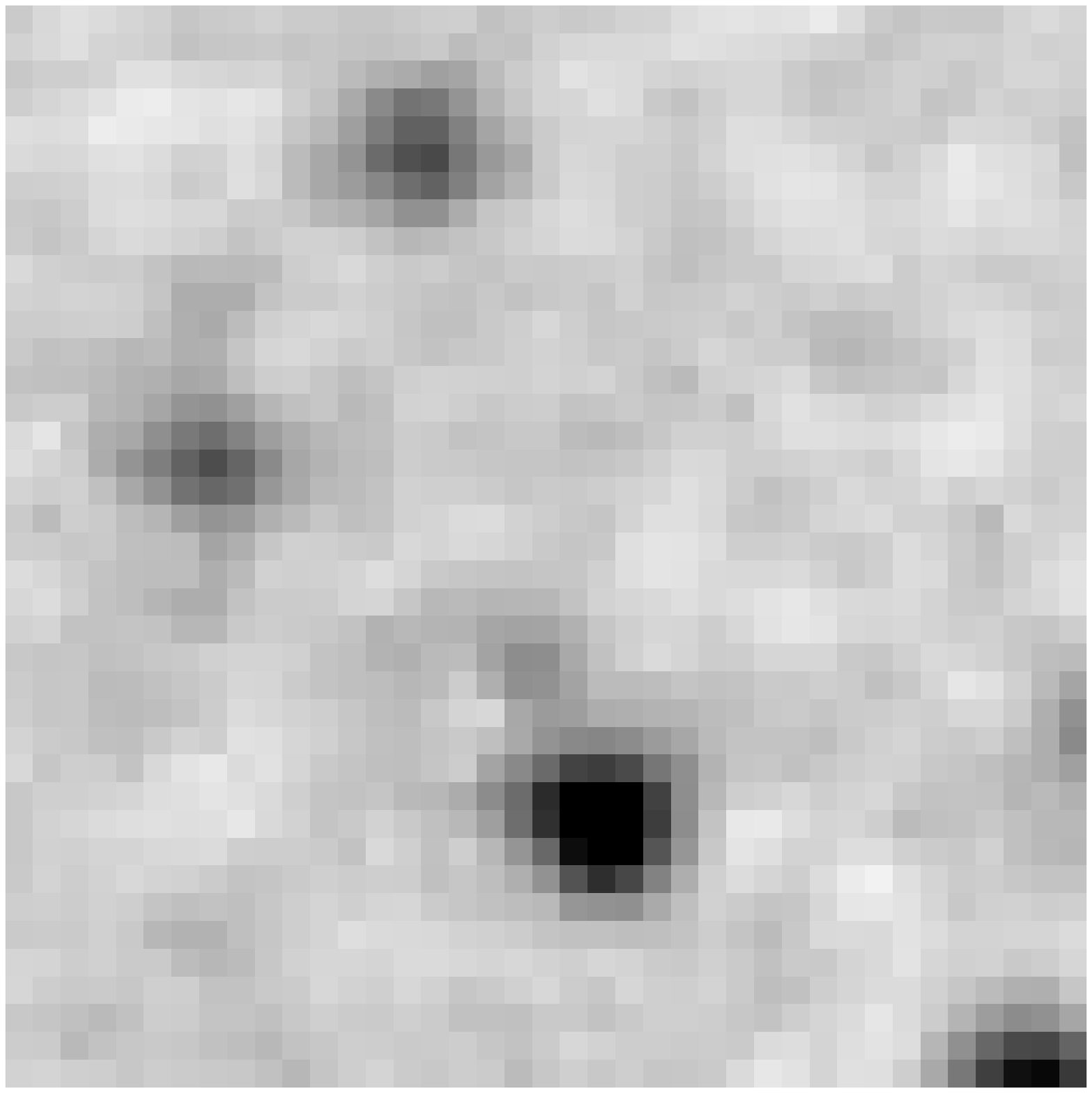}
   \includegraphics{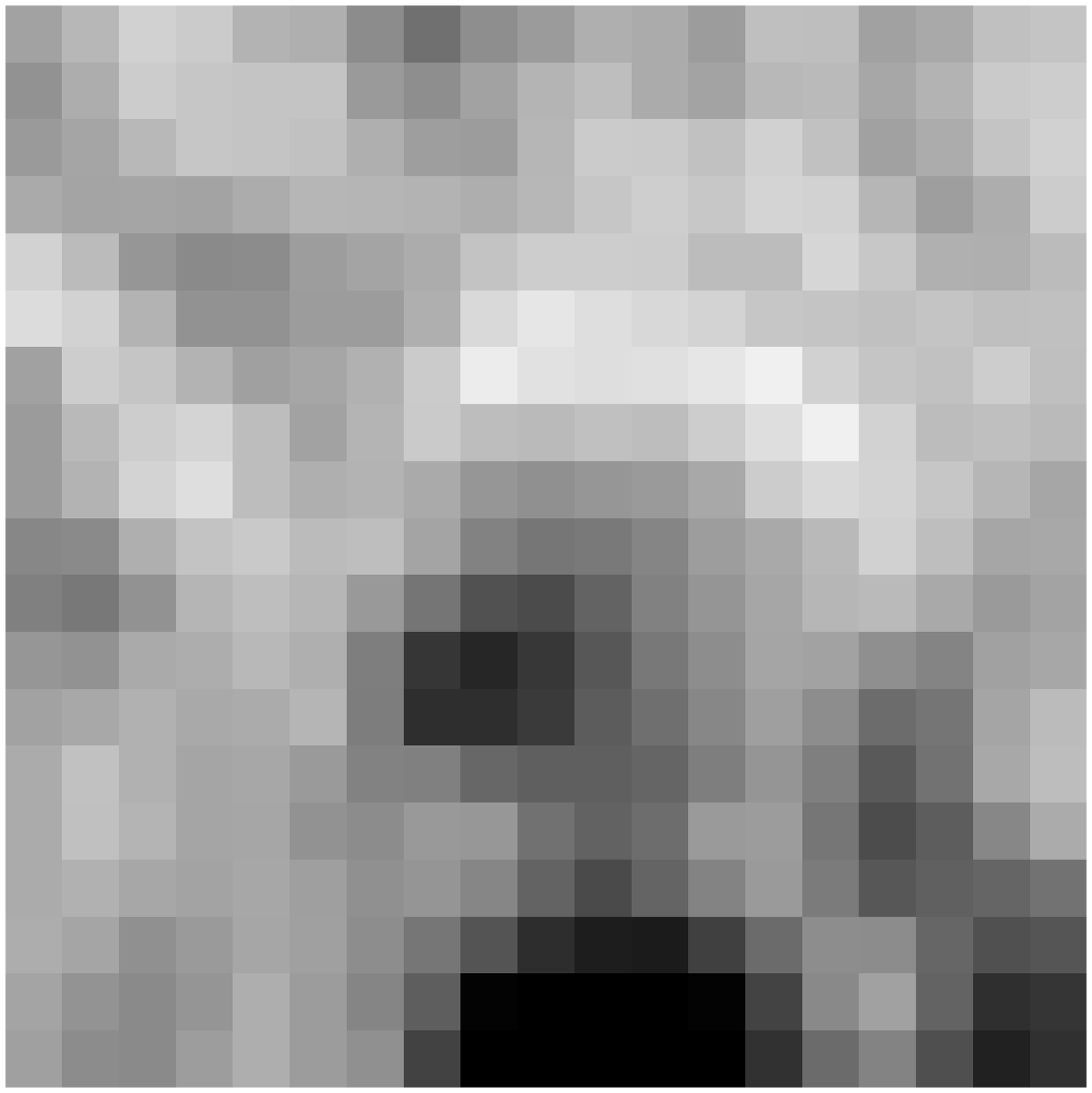}

\caption{Two examples of the power of combining high-resolution and
low-resolution radio imaging (with the VLA at 1.4 GHz) with Spitzer
IRAC $4.5\mu$m and MIPS $24\mu$m imaging to determine the galaxy
counterparts of SHADES sub-mm sources. The upper row of panels are $20
\times 20$ arcsec$^2$ postage stamp images extracted around the
position of SHADES source Lockman 850.2 from the VLA A+B+C array map,
the VLA B+C array map, the Spitzer IRAC $4.5\mu$m map, and the Spitzer
MIPS $24\mu$m map. The circle on the radio images has a radius of 8
arcsec, which is the typical search radius for identifying the
counterpart of a SHADES source with 95\% confidence. For this source,
the VLA imaging reveals two potential counterparts, but the Spitzer
imaging reveals that it is the southerly candidate which is dusty. The
second row of images shows the same information for SHADES source
Lockman 850.5. In this example there is no clear radio counterpart in
the high-resolution radio map, but the tentative radio counterpart
provided by the lower-resolution radio imaging is confirmed as real by
the Spitzer data. The radio data will be published in full by Ivison
et al. (2005). The Spitzer data for the Lockman Hole sources have been
provided to the SHADES consortium by Eichii Egami and George Rieke.}
\end{figure}

\section{VLA and Spitzer identifications}

In Fig. 2 we show 20 example $20 \times 20$ arcsec$^2$ postage stamps
extracted from our deep VLA 1.4 GHz images of the SHADES fields,
centred on the positions of the SCUBA sources. Contours from the radio
images are shown at $2, 3, 4, 5$ and $10 \sigma$. The circles have a
radius of 8 arcsec, which represents an appropriate search radius for
radio counterparts, given the uncertainties in the SCUBA
positions. What is striking about this plot is that $\simeq 15$ (i.e.,
75\%) of these sources have radio counterparts. This figure is
significantly higher than found from the radio follow-up of any
previous SCUBA survey. Such a high radio-identification rate confirms
the reality of the vast majority of the $850\mu$m sources. The fact
that the figure is so high will also be in part due to the depth and
quality of the radio data, but nevertheless it already seems clear
that few of the SHADES sources can lie at redshifts significantly in
excess of $z \simeq 3$. A corollary to this is that we can now
realistically expect to obtain an accurate ($\simeq 1$ arcsec)
position for the vast majority of the SHADES sources, providing
excellent prospects for subsequent unambiguous optical/IR
identification, and optical/infrared spectroscopy. Moreover, for those
sources which escape optical spectroscopy, the existence of a radio
detection will still assist enormously in the derivation of accurate
redshift estimates, especially in combination with BLAST and Spitzer
data.

Even the high success rate of radio identification shown in Fig. 3 is
not the whole story. The production of lower-resolution B+C array VLA
maps (i.e., deliberately leaving out the A-array data) has revealed
that even when, at first sight, a source appears to lack a radio
identification, often an extended radio counterpart is found to exist
in the lower-resolution map (resolved out in the A-array data). An
example of this is shown in Fig.  3.

Also illustrated in Fig. 3 is the added value of combining the radio
data with Spitzer imaging. Spitzer data for SHADES is being provided
for the Lockman Hole field in collaboration with the Spitzer GTO
consortium, and for the SXDF as part of the SWIRE survey. Figure 3
shows how these data can help both to differentiate between
alternative radio counterparts, and to confirm the reality and dusty
nature of tentative radio identifications.

\section{Forthcoming papers}

Detailed predictions of the extent to which SHADES can constrain the
redshift distribution and clustering of submm sources are presented by
van Kampen et al. (2004), while an overview of the survey strategy and
design is provided in Mortier et al. (2004).  In addition, three
journal papers based on the current interim dataset are currently in
preparation: Scott et al. (2005) will present the current sub-mm maps,
source list and number counts, Ivison et al.  (2005) will present
radio and Spitzer identifications, and Aretxaga et al.  (2005) will
report the estimated redshift distribution of the current SHADES
sample. This first set of data papers will be followed by publication
of the full multi-frequency study of SHADES sources, and by a detailed
comparison of our results with the predictions of current models of
galaxy formation.

\begin{acknowledgments}

I thank the members of the SHADES consortium, and also the staff of
the JCMT, for their role in turning SHADES from a simulation into a
real data set.

\end{acknowledgments}

\end{document}